\documentclass[useAMS]{mn2e}
\usepackage{psfig}
 \usepackage{url}
\usepackage{times}
\def\spose#1{\hbox to 0pt{#1\hss}}
\newcommand\lsim{\mathrel{\spose{\lower 3pt\hbox{$\mathchar"218$}}
     \raise 2.0pt\hbox{$\mathchar"13C$}}}
\newcommand\gsim{\mathrel{\spose{\lower 3pt\hbox{$\mathchar"218$}}
     \raise 2.0pt\hbox{$\mathchar"13E$}}}
\def\ltsima{$\; \buildrel < \over \sim \;$}
\def\lsim{\lower.5ex\hbox{\ltsima}}
\def\gtsima{$\; \buildrel > \over \sim \;$}
\def\gsim{\lower.5ex\hbox{\gtsima}}

\title[Radio--loud AGN and the CMB]
{Radio--loud active galactic nuclei at high redshifts and the cosmic microwave background}
\author[G. Ghisellini et al.]
{G. Ghisellini$^{1}$ \thanks{E--mail: gabriele.ghisellini@brera.inaf.it},
A. Celotti$^{2,3,1}$, F. Tavecchio$^1$, F. Haardt$^{4,5}$, T. Sbarrato$^{4,1,6}$  \\
$^1$ INAF -- Osservatorio Astronomico di Brera, via E. Bianchi 46, I--23807 Merate, Italy \\
$^2$ SISSA, via Bonomea 265, I--34135 Trieste, Italy \\
$^3$ INFN -- Sezione di Trieste, via Valerio 2, I--34127 Trieste, Italy \\
$^4$ DiSAT, Universit\`a degli Studi dell'Insubria, Via Valleggio 11, I--22100 Como, Italy \\
$^5$ INFN, Milano--Bicocca, Piazza della Scienza 3, I--20126 Milano, Italy \\
$^6$ European Southern Observatory, Karl Schwarzschild--Str. 2, D-85748 Garching bei M\"unchen, Germany \\
}

\begin{document}

\pagerange{\pageref{firstpage}--\pageref{lastpage}} \pubyear{2012}

\maketitle
\label{firstpage}

\begin{abstract}
The interaction between the emitting electrons and the Cosmic Microwave 
Background (CMB) affects the observable properties of radio--loud 
Active Galactic Nuclei (AGN) at early epochs. 
At high redshifts $z$, the CMB energy density [$U_{\rm CMB} \propto (1+z)^4$] 
can exceed the magnetic one ($U_{\rm B}$) in the lobes of radio--loud AGN.
In this case the relativistic electrons cool preferentially by  
scattering off CMB photons, rather than by synchrotron emission. 
This makes more distant sources less luminous in radio and more luminous in X--rays 
than their closer counterparts.
In contrast, in the inner jet and the hot spots, where $U_{\rm B}>U_{\rm CMB}$,
synchrotron radiation is unaffected by the presence of the CMB. 
The decrease in radio luminosity is thus more severe in 
misaligned (with respect to our line of sight) high--$z$ sources,
whose radio flux is dominated by the extended isotropic component. 
These sources can fail detection in current flux limited radio surveys,
where they are possibly under--represented. 
As the cooling time is longer for lower energy electrons, the radio 
luminosity deficit due to the CMB is less important at low radio frequencies. 
Therefore objects not detected so far at a few GHz 
could be picked up by low frequency deep surveys, such as LOFAR and SKA. 
Until then, we can estimate the number of high--$z$ radio--loud AGN through the census of 
their aligned proxies, i.e., blazars, since their observed radio emission arises in 
the inner and strongly magnetized compact core of the jet, and it is 
not affected by inverse Compton scattering off CMB photons.
\end{abstract}

\begin{keywords}
galaxies: BL Lacertae objects: general --- galaxies: quasars: general  --- 
radiation mechanisms: non-thermal --- radio continuum: general  --- 
cosmology: cosmic microwave background  
\end{keywords}


\section{Introduction}

Radio emission from jetted extragalactic sources associated with AGN 
is produced in two main  distinct locations, i.e., the relativistic jet and the 
at most mildly relativistic
radio lobes (including the hot spots), where jet material impacts on to
the intergalactic and intracluster media, and  its mechanical power 
is partly dissipated in a shock (e.g. Begelman et al. 1984, De Young 2002).
Radiation is strongly enhanced at viewing angles close to the jet axis by relativistic beaming, 
and it becomes very faint, until undetectable, for large
viewing angles.
On the contrary, the slower expanding extended lobes 
emit almost isotropically. 
The  lobes are usually detected in the radio band thanks to their optically thin synchrotron emission, 
which is characterized by a relatively steep energy spectral index,  
$\alpha\gsim$0.7 [$F(\nu) \propto \nu^{-\alpha}$]. 
The radio jet emission, instead, is modelled as the superposition of several
(partially) self--absorbed synchrotron components, produced by more compact structures, 
giving rise to a flatter radio spectrum ($\alpha\lsim$0.7).

The idea of unification of all radio--sources 
on the basis of our viewing angle dates back to Blandford \& Rees (1978; see Urry \& Padovani 1995 for a review). 
When our line of sight lies within an angle $\sim 1/\Gamma$ to the jet axis, where $\Gamma$ 
is the jet bulk Lorentz factor, the beamed, flat radio flux dominates, and the source 
is classified as a blazar. For larger viewing angles isotropic emission 
from the extended regions takes over, and we do observe a steep spectrum radio source. 
The ``$1/\Gamma$" beaming angle provides an approximated but convenient 
way to relate the number density of the observed blazars to that of the underlying parent population. 
Indeed, as discussed in Volonteri et al. (2011; see also Ghisellini et al. 2013), 
for each observed blazar there must exist a number $N \simeq 2\Gamma^2$ of sources intrinsically 
similar, but pointing away from us. The spectrum of such population should be steep in the radio band, 
while at larger frequencies isotropic components (e.g., emission lines, accretion disk, host galaxy), 
usually -- but not always -- masked in blazars by the  relativistically boosted jet emission, dominate.

Blazars are easily detectable up to high redshifts. 
Although relatively rare, their large fluxes in almost all electromagnetic bands 
make them detected in large area flux limited surveys. 
The spectral energy distribution (SED) of blazars is characterized by two
broad humps, usually interpreted as the synchrotron and the inverse
Compton emissions from a single 
distribution of highly relativistic electrons (Sikora, Begelman \& Rees 1994; Ghisellini et al. 1998). 
At increasing  bolometric luminosities, the peak of the two humps shift to lower frequencies, 
and the high energy hump becomes comparatively stronger (Fossati et al. 1998; Donato et al. 2001).
This phenomenology suggests that hard X--ray and $\gamma$--ray surveys are 
particularly efficient in finding powerful blazars at high redshifts.
Indeed, such expectations are confirmed by the all sky surveys  
carried out by the Burst Alert Telescope (BAT) onboard {\it Swift} (detecting
blazars up to $z\sim 4$, see Ajello et al. 2009), and
by the large Area Telescope (LAT) onboard {\it Fermi} 
(detecting blazars up to $z\sim 3$, see e.g. Nolan et al. 2012).

In this work we will consider the subclass of powerful radio--sources, classified as FR II 
radio galaxies {(Fanaroff \& Riley 1974)} and steep spectrum radio--loud quasars, where the 
relativistic jet is misaligned, and flat spectrum radio quasars (FSRQs), whose jet points at us. 
These AGNs are the most powerful persistent sources we know of, 
their jets carrying a power, in magnetic fields and particles, that can exceed
$10^{47}$ erg s$^{-1}$ (Celotti \& Ghisellini 2008; Ghisellini 2010; Fernandes et al. 2011).  
Estimates of the energy associated to large scale extended structures yield to 
at least $10^{60}$--$10^{61}$ erg, shared amongst relativistic emitting electrons, 
magnetic fields, and  possibly protons of still unknown temperature 
(e.g. Blundell \& Rawlings 2000; Croston et al. 2005; Belsole et al. 2007; Godfrey \& Shabala 2013).

In powerful blazars, the synchrotron hump peaks at sub--mm frequencies, with a steep 
spectrum above. 
The rapidly declining non--thermal emission ($\alpha\gsim 1$) allows us 
to detect the accretion disc component. 
Fits with a standard Shakura \& Sunyaev (1973)
disc emission model (Calderone et al. 2013; Castignani et al. 2013)
imply for all powerful blazars at $z>2$ in the 
BAT and LAT catalogs, black hole masses in excess of $10^9 M_\odot$ 
(Ghisellini et al. 2010a; 2011).
This fact boosted interest in the search of  powerful blazars at high--$z$, as an efficient way
to count supermassive black holes at early epochs. 
In this context, blazars counts can be competitive with (mainly optical)
searches for mostly radio--quiet quasars (i.e. the Sloan
Digital Sky Survey, SDSS; York et al. 2000).

As already mentioned, for each detected blazar, there must exist other 
$\sim 2\Gamma^2$ misaligned sources with similar properties, including $M$. 
Typical values of $\Gamma$ are between 10 and 15, so the very presence of a single 
blazar with a black hole mass $M >10^9 M_\odot$ already 
implies the existence of other $450 (\Gamma/15)^2$ sources with similar masses.
Should such sources have extended structures similar to those seen at lower
redshifts, they could be easily detected -- even at $z>3$ -- 
in radio surveys like the FIRST (Faint Images of the Radio Sky at 20 cm;
White et al. 1997), characterized by a flux limit of 1 mJy at 1.4 GHz.
Instead these sources appear to be at best under--represented in the combined SDSS+FIRST survey 
(see Volonteri et al. 2011). 
The chain of arguments leading to the claimed deficit was the following:
first, we established that all BAT blazars with $z>2$ and
an X--ray luminosity (in the 15--55 keV band) $L_X>2\times 10^{47}$ erg s$^{-1}$
have $M > 10^9 M_\odot$.
Then, we exploited the luminosity function of Ajello et al. (2009)
to find out the expected density of blazars of high 
$L_X$ ($\gsim 10^{47}$ erg s$^{-1}$, presumably hosting a black hole with 
$M\gsim 10^9 M_\odot$) in different redshift bins. 
To be conservative, we modified the Ajello et al. (2009) luminosity 
function above $z=4$, introducing an exponential cut--off  at larger redshifts.
Adopting $\Gamma=15$ as an average value (estimated by fitting the SED
of all the BAT blazars with $z>2$, Ghisellini et al. 2010a, 2010b), we then computed the overall density
of powerful radio--loud AGNs hosting a $M>10^9M_\odot$ black hole, in different redshifts bins.
Finally we compared this expected number density with  the detections in  the SDSS+FIRST survey,
finding agreement at $z\lsim 3$, and a deficit of at least a factor of $10$ above.

Recently we showed that the source B2 1023+25 at $z=5.3$ is a blazar (Sbarrato et al. 2012; 2013).
This object, present in the SDSS+FIRST survey,
was studied by Shen et al. (2011), and is one of the only 4 
sources detected in radio above $z>5$ in the Shen et al. (2011) catalog.
For B2 1023+25 we estimated $M=(3\pm 1.5)\times 10^9 M_\odot$, and
a bulk Lorentz factor $\Gamma=13$.
If its  viewing angle is indeed $< 1/\Gamma$,  we expect other $\sim$340$(\Gamma/13)^2$ 
radio--sources in the SDSS+FIRST survey, while there are only 3.
Given the importance of verifying the true number density
of massive black holes in the early Universe, we are urged
to explain such a discrepancy. 

Volonteri et al. (2011) discussed this issue, suggesting
three possible causes: 
i) the radio lobes become for some (yet unknown) reason weaker
than expected at $z \gsim 3$, falling below the FIRST detection threshold;
ii) the average $\Gamma$ is typically(again for unknown reasons) smaller than 10--15
at high redshifts, so that the factor $2\Gamma^2$ is much smaller than  assumed; 
iii) powerful misaligned radio sources at high redshifts do not enter 
the SDSS survey because of heavy obscuration in the optical, so that they
lie below its detection threshold; 
iv) missing identifications of extended radio structures associated to lobes of an AGN.

Whatever the reason is, we face  the intriguing fact that ``something" must change for sources at  $z\gsim$3.   
This led us to assess the effects of the Cosmic Microwave Background 
(CMB) on the radio properties of the extended structures of powerful sources, 
and here we present our findings. 
The idea of the CMB affecting the behaviour of radio loud AGN dates back to the works by Felten \& Rees (1967) 
and Rees \& Setti (1968), and it was followed by Krolik \& Chen (1991) to explain the prevalence of sources 
with steep radio spectra at high redshifts. 
Later Celotti \& Fabian (2004) demonstrated that the extended 
structures of high--$z$ powerful radio sources could contribute significantly to the diffuse X--ray 
background, via inverse Compton (IC) scattering between radio emitting electrons and CMB photons, 
while more recently Mocz, Fabian \& Blundell (2011) studied the evolution of the extended structure 
of a radio source calculating synchrotron and adiabatic losses, as well as IC losses on CMB photons, 
whose radiation energy density is larger at larger redshifts.

Here, we concentrate on the synchrotron emission from those relativistic electrons 
cooling mainly by scattering off CMB seed photons.
Since the energy density of the CMB scales as $(1+z)^4$,
it may become larger than the typical magnetic energy density
in radio lobes, enhancing the total radiative losses, and hence modifying the electron distribution function, 
especially at the largest energies. We will compute in detail how the CMB modifies the electron distribution 
in jets, hot spots and extended lobes, and how the radio emission from powerful radio--loud AGNs changes 
with redshift, allowing us to compare our results with observations. 

In this work, we adopt a flat cosmology with $H_0=70$ km s$^{-1}$ Mpc$^{-1}$ and
$\Omega_{\rm M}=0.3$. 

\section{Radio galaxies and the CMB}
\subsection{The environment of radio galaxies at different epochs}

At all redshifts, the extended structure of a radio source is due to the energy 
deposited by the jet once it impacts onto the intergalactic medium. 
Today, powerful radio--sources 
can  easily  reach $\sim$1 Mpc away from the central engine.
These radio sources are often associated with poor clusters or groups, and 1 Mpc  
is larger than the core radius $r_{\rm c}\sim 0.1$ Mpc of a typical cluster, 
where the density is estimated to be about $n_{\rm c}\sim 10^{-3}$ cm$^{-3}$ (e.g. Sarazin 2008). 
Out of the core radius the gas density approximatively follows an isothermal profile,  
$n = n_{\rm c}/[ 1+(r/r_{\rm c})^2]$, resulting in $n \sim 10^{-5}$ cm$^{-3}$ at $r \sim 1$ Mpc. 
Such density is $\sim$30 times larger than the mean cosmic density of baryons today.

We do not know the typical sizes, masses and gas densities of {\it proto}--clusters that 
can host powerful radio sources at $z\gsim 3$. 
What we do know is that the large black holes ($M \gsim 10^9 M_\odot$) 
required some time to accrete their masses. 
As an example, assuming a Salpeter time of $40$ Myr, the time needed by a $10^2 M_\odot$ 
black hole seed to reach $10^9 M_\odot$ is $\simeq 930$ Myr.  
If the seed mass is 
one hundred times larger, the time required is still $\simeq 660$ Myr. 

Such timescales are  long enough for an early AGN to fully develop the extended radio lobes. 
In fact, the far end of the jet, assumed to be advancing at $\sim 10\%$ of the light speed,
takes only 32 Myr to reach a distance of $\sim$1 Mpc.  
As the proto--cluster may have not completely formed yet at such high $z$, 
it is likely that the density at 1 Mpc from the center is similar to the
average cosmic baryon density, i.e. $n_{\rm Mpc} \sim 2\times 10^{-5} [(1+z)/4]^3$ cm$^{-3}$.

The above simple estimates show our point:  
at high redshifts the medium at $\sim 1$ Mpc from a radio--loud AGN 
(where the
jet dissipates its own energy) has a 
density that, by chance, is similar to the density of the intracluster medium at $r \sim 1$ Mpc today. 
As a consequence, radio lobes of similar power should have
similar sizes.
If the magnetic field is always around equipartition, it should  have a similar value
today and at $z\simeq 3$. 

We can now study the physics of the interaction between the extended emission 
of a jet and the CMB in details.

\subsection{Cooling rates}

As the energy density of the cosmic microwave background is
\begin{equation}
U_{\rm CMB}\, =\,  aT_0^4 (1+z)^4 \, =4.22\times 10^{-13} (1+z)^4 \, \,
{\rm erg \, cm^{-3}},
\end{equation}
where $T_0=2.725$ K is the CMB temperature today, 
the value of a magnetic field in equipartition with $U_{\rm CMB}$ is
\begin{equation}
B_{\rm CMB}\, =\,  [8\pi U_{\rm CMB}]^{1/2} \, =\, 3.26\times 10^{-6} (1+z)^2 \,\, {\rm G}.
\label{beq}
\end{equation}
At $z=3$, we have $B_{\rm CMB}$=52 $\mu$G, and $B_{\rm CMB}$=120 $\mu$G at $z=5$.

Relativistic electrons contained in a plasma where $B<B_{\rm CMB}$
would preferentially cool down via IC, rather than synchrotron losses. 
Consequently the radio emission is expected to be quenched, or severely weakened. 
In order to quantify the strength of such effect, we need first 
to consistently determine the particle energy distribution. 
We will do this under the simplifying hypothesis of continuous injection of 
fresh particles, and we will account for radiative and adiabatic cooling.

A further approximation is related to the fact that 
particles are likely injected (i.e., accelerated to relativistic energies) by shocks in the hot spots, 
and from there they diffuse out to fill up the extended radio lobes. 
The particle energy distribution is therefore a function of the distance
from the injection region. For simplicity, we evaluate instead an average distribution, adopting a 
spatially--averaged magnetic field. This can be thought as equivalent to observe the radio 
source with poor angular resolution. 

With the above simplifying assumptions, the adiabatic and radiative cooling rates can be written as
\begin{equation}
\dot \gamma_{\rm ad}\, =\, {\gamma  v_{\rm exp}  \over R}
\, \equiv \, {\gamma  \beta_{\rm exp}  \over R/c},
\end{equation}
\begin{equation}
\dot \gamma_{\rm rad} \, =\, {4\over 3} \,
{\sigma_{\rm T} c \gamma^2\over m_{\rm e}c^2} ( U_{\rm B} + U_{\rm CMB} + U_{\rm s} ),
\end{equation}
respectively,  where $v_{\rm exp}$ is the expansion speed of the lobe, 
and $U_{\rm s}$ is the energy density of synchrotron radiation 
as electrons cool also via the synchrotron self--Compton (SSC) mechanism.
For the conditions considered here these latter radiative losses are 
significantly smaller than synchrotron and IC
ones off CMB photons, and can be thus safely neglected. 

For a given extended region of length--scale $R$, adiabatic losses  exceed the radiative 
ones for particle Lorentz factors up to a value $\gamma_{\rm ad}$ given by
\begin{eqnarray}
\gamma_{\rm ad}\, &=&\,  {3 m_{\rm e}c^2 \beta_{\rm exp}  \over 
4 R \sigma_{\rm T}    ( U_{\rm B} + U_{\rm CMB} ) } 
\nonumber \\
&\approx & \, { 7.2\times 10^{5} \beta_{\rm exp,-1}  \over 
 (1+z)^4 R_2 ( 1+ U_{\rm B}/U_{\rm CMB}) }.
\end{eqnarray}
Here $R_2=R/100$ kpc, and $\beta_{\rm exp,-1}=\beta_{\rm exp}/0.1$. 

Adiabatic losses ($\propto \gamma$) do not alter the shape of the 
injected electron distribution, while 
on the contrary radiative losses ($\propto \gamma^2$) steepen it. 
In the presence of a constant injection rate of a power--law distribution, 
\begin{equation}
Q(\gamma)\propto \gamma^{-s},  
\end{equation}
the slope of the evolved electron distribution $N(\gamma)$ steepens by one unity for Lorentz factors 
$\gamma > \gamma_{\rm rad}$, as long as the condition $\gamma_{\rm rad} > \gamma_{\rm ad}$ holds.
The value of $\gamma_{\rm rad}$ is given by
\begin{equation}
\gamma_{\rm  rad} \, =\, { 3 m_{\rm e}c^2 \over 
4\sigma_{\rm T} c \Delta t\, U_{\rm CMB} (1+U_{\rm B}/U_{\rm CMB}) },
\end{equation}
where $\Delta t$ is the time elapsed since a particle is injected in the source 
and can be regarded as the time spent by the particle inside the extended region. 

In the present context, the region itself could consist of many smaller sub--regions 
whose length--scale is of the order of the coherence length of the magnetic field, $\lambda$. 
Particles then follow a sort of random walk with 
mean free path $\approx \lambda$, before escaping the source. We can thus write  
\begin{equation}
\Delta t \, \approx \, {R\over c} \left(1+ {R \over \lambda} \right)
\label{deltat}
\end{equation}
and, in turn, 
\begin{equation}
\gamma_{\rm  rad} \, \approx \, { 4.9\times 10^6 \over 
(1+z)^4 R_2 [(1 + R/\lambda)] (1+U_{\rm B}/U_{\rm CMB}) }.
\end{equation}
Cooling electrons emit at synchrotron frequency 
\begin{eqnarray}
\nu_{\rm s, cool} &\approx & 3.6\times 10^6 B \gamma_{\rm  rad}^2 \,\,\,\,  {\rm Hz}
\nonumber \\
&=& { 1.8\times 10^{15} B_{-5}  \over 
\left\{(1+z)^4 R_2 [(1 + R/\lambda)] (1+U_{\rm B}/U_{\rm CMB})\right\}^2 }.
\end{eqnarray}
If we assume $\lambda\sim 10$ kpc (Celotti \& Fabian 2004; Carilli \& Taylor 2002), 
we find that, for $B=10$ $\mu$G, the main radiative coolant is synchrotron emission for $z\lsim 1$,
while at higher redshifts IC off CMB photons dominates (see Eq. \ref{beq}).
For reference, at $z=2$ ($z=3$) the rest frame cooling frequency 
$\nu_s$ is 2 GHz (0.2 GHz). 

\subsection{Relativistic energy distribution of emitting particles}

The time--dependent relativistic energy distribution $N(\gamma,t)$ (in cm$^{-3}$)   
of the emitting particles is calculated
through a continuity equation, for a constant injection rate of fresh particles
at rate $Q(\gamma)$ (in cm$^{-3}$ s$^{-1}$),
and assuming adiabatic and radiative cooling. 
Electrons emit synchrotron photons in the radio band, and IC in X/$\gamma$--rays.   

The solution of the continuity equation is time dependent. 
For illustrative purposes, in the following we will show results 
computed at time $t=\Delta t$ after the start of particle injection. 
We remind that $\Delta t$ is the particle escape time from the source. 
Such characteristic time is approximatively the time needed for the particle 
distribution to reach a stationary configuration.
This is strictly true in the case of steady continuous injection lasting a time 
$\gg \Delta t$. 
We must note however that even in this case the particle distribution may never 
reach a stationary state, since source expansion most likely decreases the average magnetic field
(see e.g. Mocz, Fabian \& Blundell 2011).
With such caveat in mind, we can  now compute particle distributions and the relative emission spectra. 
To do so, we further assume that the magnetic field is constant for a time $\Delta t$. 
At  high energies, $\gamma \gsim \gamma_{\rm  rad}$, we have 
\begin{equation}
N(\gamma)\, \sim \,  \int_\gamma  {Q(\gamma^\prime) \over \dot \gamma_{\rm rad}} \, d\gamma^\prime
\end{equation}
where $N(\gamma)\equiv N(\gamma,\Delta t)$.  
As long as $U_{\rm B}\gg U_{\rm CMB}$, increasing $U_{\rm CMB}$ leaves
$N(\gamma)$ unaltered. The high energy emission increases, but the radio synchrotron
flux stays constant. On the contrary, when $U_{\rm CMB}$ exceeds $U_{\rm B}$, $N(\gamma)$ is bound to decrease.
This implies a constant high energy flux -- since the increase of $U_{\rm CMB}$
exactly compensate for the decrease of $N(\gamma)$ -- while the  (radio) synchrotron flux
{\it decreases}.

At lower energies, $\gamma \lsim \gamma_{\rm  rad}$, radiative losses are negligible and hence 
$N(\gamma)$ has the very same slope of the injected distribution $Q(\gamma)$, 
and a normalization set by adiabatic losses and particle escape. 
If adiabatic losses prevail, then 
\begin{equation}
N(\gamma)\sim \int_\gamma {Q(\gamma^\prime) \over \dot \gamma_{\rm ad}}\, d\gamma^\prime 
\end{equation}
if instead escape dominates, then 
\begin{equation}
N(\gamma)\sim \Delta t \, Q(\gamma).
\end{equation}
As long as adiabatic cooling dominates, $N(\gamma)$ becomes insensitive to changes of $U_{\rm CMB}$.
This means that at low radio frequencies the same source will have the same luminosity
for any $U_{\rm CMB}$, i.e., independent of redshift.
It should be noted, however, that this could happen at electron energies corresponding
to the self--absorbed part of the synchrotron spectrum, and therefore the effect could be difficult
to observe, unless $\gamma_{\rm  rad}$ is significantly larger than the self--absorption energy.
This latter circumstance may happen if the escape time is short, i.e., for a small source size.

In the following section we will show a few examples of solutions of the continuity equation discussed above.

\begin{figure}
\vskip -0.6 cm
\hskip -2cm
\psfig{file=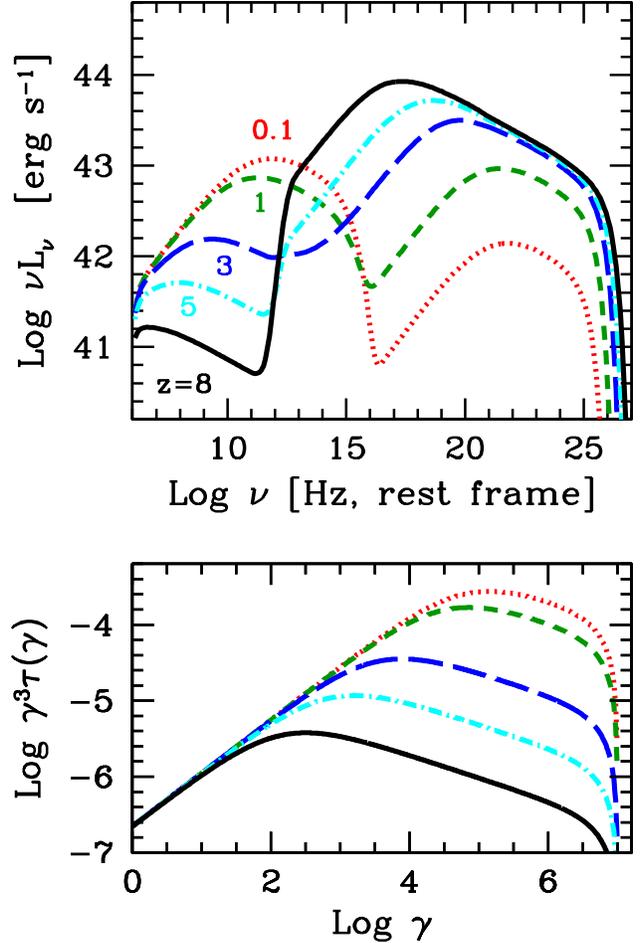,height=14.cm,width=14.5cm}
\vskip -1 cm
\caption{Top panel: SEDs in the rest frame for different redshifts,
as labeled. 
The size of the source is 100 kpc.
Particles are injected with a single power--law distribution
($s=2.3$) and the magnetic field is $B=10\mu$G.
Bottom panel: electron energy distribution, plotted in the
form $\gamma^3\tau(\gamma)$ as a function of $\gamma$,
where $\tau(\gamma) \equiv \sigma_{\rm T} R N(\gamma)$.
In this representation, the peak of $\gamma^3\tau(\gamma)$
corresponds to electrons
emitting at the peak of the SED.
} 
\label{lobozpl}
\end{figure}
\begin{figure}
\vskip -0.5cm
\hskip -0.1cm 
\psfig{file=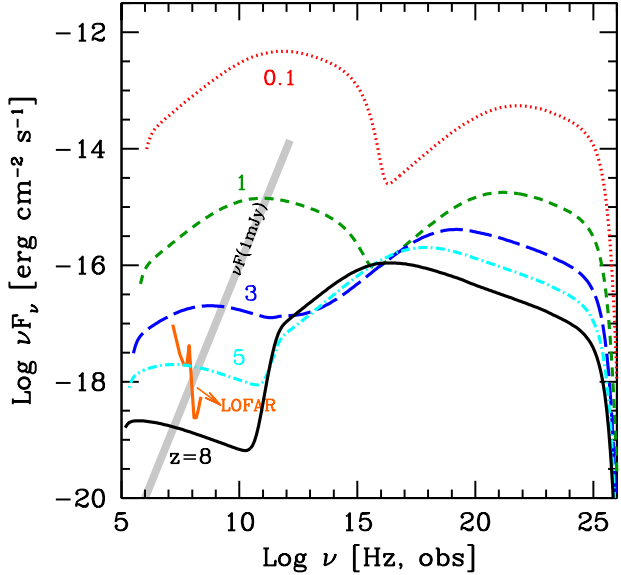,height=9.cm,width=9.cm}
\vskip -0.5 cm
\caption{The same  as in  Fig. \ref{lobozpl},  but in  the observer frame.
For illustration, we show the line corresponding to $F(\nu)=1$ mJy
and the limiting sensitivity of LOFAR, once completed. 
The FIRST survey, corresponding to 1 mJy at the observed frequency 
of 1.4 GHz, can detect the source up to $z\sim 3$, but not beyond.
} 
\label{lobozplf}
\end{figure}

\subsection{Examples of energy distributions}

We assume our idealized source as homogeneous and spherical,
with a homogeneous and tangled magnetic field filling it up. 
To highlight the role of the CMB in determining the observable properties of radio--loud AGNs, 
we assume that the structure and the basic intrinsic
properties of the source are independent of redshift.

The size of the extended source is taken as $R=100$ kpc. 
Relativistic electrons are injected throughout the
source at constant rate, and we fix the magnetic field coherence length to 10 kpc. 
Finally, the entire source is assumed to expand at $\beta_{\rm exp}=0.1$.

\subsubsection{Single power--law injection}

We start by assuming that electrons are injected
throughout the source with a simple power law distribution, 
i.e., $Q(\gamma)\propto \gamma^{-2.3}$
between $\gamma_{\rm min}$=1 and $\gamma_{\rm max}= 10^7$
and $B=10$ $\mu$G throughout the entire region.
The assumed slope is within the range predicted by shock acceleration 
(e.g. Kirk \& Duffy 1999) and what observed in real sources (e.g. Kataoka \& Stawarz 2005).
The power injected in relativistic electrons is always
$P= 5\times 10^{45}$ erg s$^{-1}$. 
This injected power yields -- according to Eqs. (12), (13) -- to an energy content in relativistic electrons
close to equipartition with the magnetic field energy:
$\log (E_{\rm B}/{\rm erg})=59.8$ and $\log (E_{\rm e}/{\rm erg})=59.6$ (at $z=0.1$)
and $\log (E_{\rm B}/{\rm erg})=59.8$ and $\log E_{\rm e}/{\rm erg})=59.5$ (at $z=8$).
The energy content in particles is not exactly the same at different $z$, but
it decreases at large redshifts, since their cooling increases.
We compute the electron density distribution and the generated SED as  discussed above.

In the top panel of Fig. \ref{lobozpl} we show 
the predicted SED of a source of equal intrinsic properties, but
located at different redshifts. The bottom panel illustrates  the corresponding electron distributions. 
The electron distribution is plotted as $\gamma^3\tau(\gamma)=\sigma_{\rm T} R \gamma^3N(\gamma)$, 
so that its peak corresponds to the Lorentz factor value that emits at the two peaks 
(synchrotron and IC) in the $\nu L_\nu$ photon distribution. 

Fig. \ref{lobozpl} allows us to appreciate the large evolution of 
the predicted SED as the redshift increases.
Inverse Compton cooling increases as $U_{\rm CMB}\propto (1+z)^4$,
and the corresponding losses, for the assumed parameters, become equal to the synchrotron ones 
at $z \simeq 0.75$ (see Eq. \ref{beq} for $B=10$ $\mu$G).
As the redshift increases, the source becomes more and more Compton dominated, 
as was pointed out by Celotti \& Fabian (2004), who also emphasized the
diffuse nature of the produced X--ray radiation. 
Celotti \& Fabian (2004) argued that such diffuse high energy emission could represent 
a large part of the {\it truly diffuse} X--ray background.

As already mentioned, the behaviour of the synchrotron flux, that steadily decreases 
as the redshift increases is due to the increased radiative cooling that forces   
the high energy tail of $N(\gamma)$ to decrease (see the bottom panel of Fig. \ref{lobozpl}).
We also notice that, even in our simple case of a single power law injection, 
the resulting $N(\gamma)$ can be described as a broken power law,  
breaking at the energy that corresponds to the two peaks of the SED.
This corresponds to the Lorentz factor for which $\dot\gamma_{\rm ad}=\dot\gamma_{\rm rad}$
(i.e. $\gamma\sim 2.8\times 10^4$ for $z=1$,
$\gamma\sim 550$ for $z=5$, cfr Eq. \ref{beq}).
The location of the  peaks shifts to lower frequencies as the redshift 
increases, due to the increased cooling, making $N(\gamma)$ to break 
at lower energies, (see the bottom panel of Fig. \ref{lobozpl}).

\begin{figure}
\vskip -0.6cm
\hskip -2cm
\psfig{file=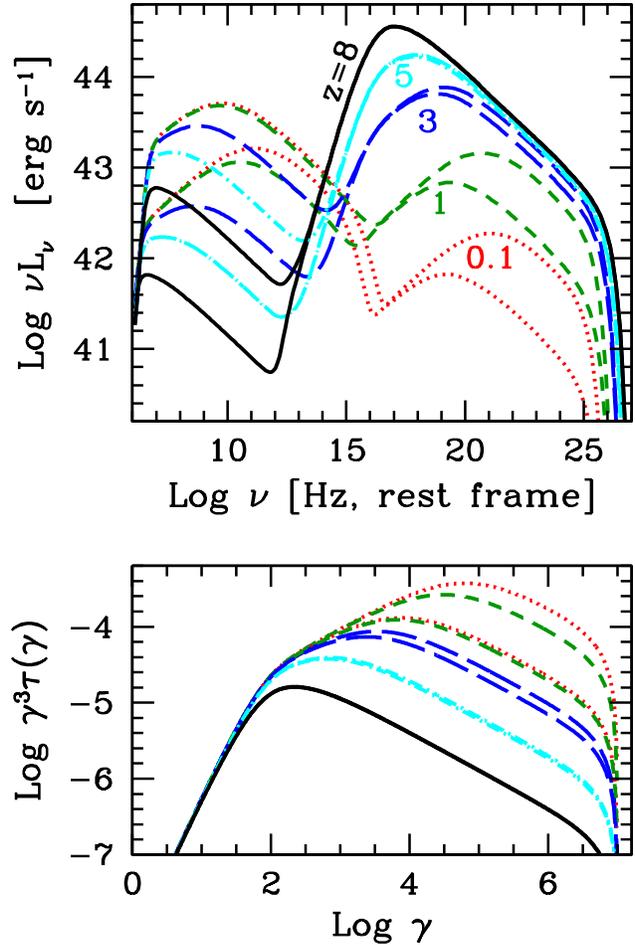,height=14.cm,width=14.5cm}
\vskip -1 cm
\caption{Top panel: SEDs in the rest frame for different redshifts,
as labeled, for a double power--law injections distribution ($s_1=0, s_2= 2.5$).
The size of the source is 100 kpc.
Bottom panel: corresponding electron energy distributions.
Solid lines: $B=10\mu$G; dashed lines: $B=30\mu$G.
} 
\label{loboz}
\end{figure}
\begin{figure}
\vskip -0.5cm
\hskip -0.1cm
\psfig{file=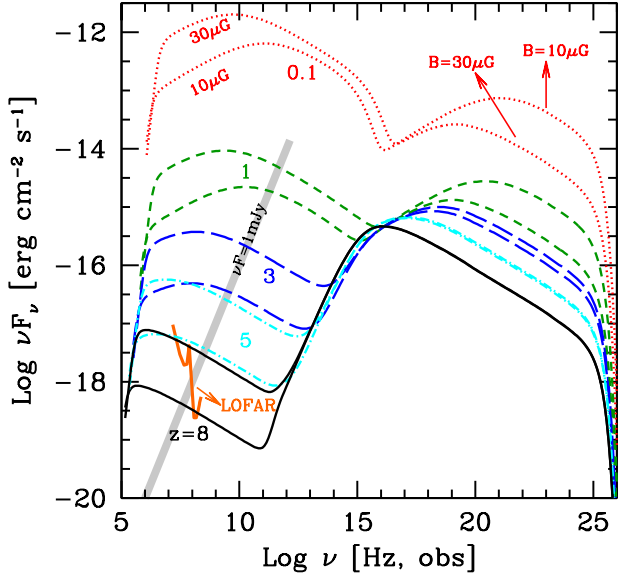,height=9.cm,width=9.cm} 
\vskip -0.5 cm
\caption{
SEDs corresponding to the cases shown in Fig. \ref{loboz}
in the observer frame.
For each redshift, two models are shown,
for a magnetic field of 10 and 30 $\mu$G. 
A greater magnetic field implies a brighter synchrotron
and a fainter IC flux.
For illustration, we show the line corresponding to 1 mJy
and the limiting sensitivity of LOFAR, once completed. 
} 
\label{lobozf}
\end{figure}

Fig. \ref{lobozplf} shows the same SEDs of Fig. \ref{lobozpl},
but in the observer frame.
We  report for comparison a limiting sensitivity of 1 mJy at all radio
frequencies, and the limiting sensitivity of the LOw--Frequency ARray
(LOFAR, see e.g. van Haarlem et al. 2013), in order to assess
the detectability of extended radio sources at high redshifts. 
Our ideal source could be detected at 1 GHz
above 1 mJy  up to  $z\sim 3$. 
To push detection further  away, we need a better sensitivity:
LOFAR can detect these sources at intermediate radio--frequencies up to
$z\sim 5$.
We remind that the plotted fluxes refer only to the {\it isotropic, extended}
emission, thus neglecting the beamed jet emission altogether.

\subsubsection{Double power--law injection}

Here we assume the same parameters as in the previous case,
but the injected particle distribution $Q(\gamma)$ is assumed to be a (smoothly 
joining) double power law:
\begin{equation}
Q(\gamma)  \, = \, Q_0\, { (\gamma/\gamma_{\rm b})^{-s_1} \over 1+
(\gamma/\gamma_{\rm b})^{-s_1+s_2} } \quad {\rm [cm^{-3} s^{-1}]}.
\label{qgamma}
\end{equation}
Fig. \ref{loboz} and Fig. \ref{lobozf} show the results
of assuming $s_1=0$, $s_2=2.5$ and a break Lorentz factor 
$\gamma_{\rm b}=100$. 
The $Q(\gamma)$ distribution
extends between $\gamma_{\rm min}\sim 1$ $\gamma_{\rm max}\sim 10^7$.
In the case of $B=10$ $\mu$G electrons and magnetic field attain nearly equipartition
at $z=0.1$, where $\log (E_{\rm B}/{\rm erg})=59.8$ and $\log (E_{\rm e}/{\rm erg})=59.7$.
At the same redshift,
for $B=30$ $\mu$G the source is instead magnetically dominated, as 
$\log (E_{\rm B}/{\rm erg})=60.75$ and $\log (E_{\rm e}/{\rm erg})=59.4$.

Fig. \ref{lobozf} shows the same SEDs of Fig. \ref{loboz}, but in the 
observed frame. 
In the $B=10$ $\mu$G case, an extended source becomes undetectable
by LOFAR in the 100 MHz -- 1 GHz band at $z \gsim 5$. 
The general behavior is similar to the single power law case:
the Compton dominance increases with $z$, and at the same time
 the  peaks shift to  lower frequencies due to the increase in Compton cooling. 

At low redshifts, $z\lsim1$, the cooling frequency exceeds  
the break frequency $\nu_{\rm b}$ (corresponding to $\gamma_{\rm b}$, Eq. \ref{qgamma}), 
and the SED shows two  distinct breaks in the synchrotron dominated range. 
The cooling frequency becomes smaller than $\nu_{\rm b}$ at higher redshifts,  
and a single break, corresponding to the $\nu F_\nu$ peak of
the synchrotron emission, is hence featured.

\begin{figure}
\vskip -0.5cm
\hskip -2.7cm
\psfig{file=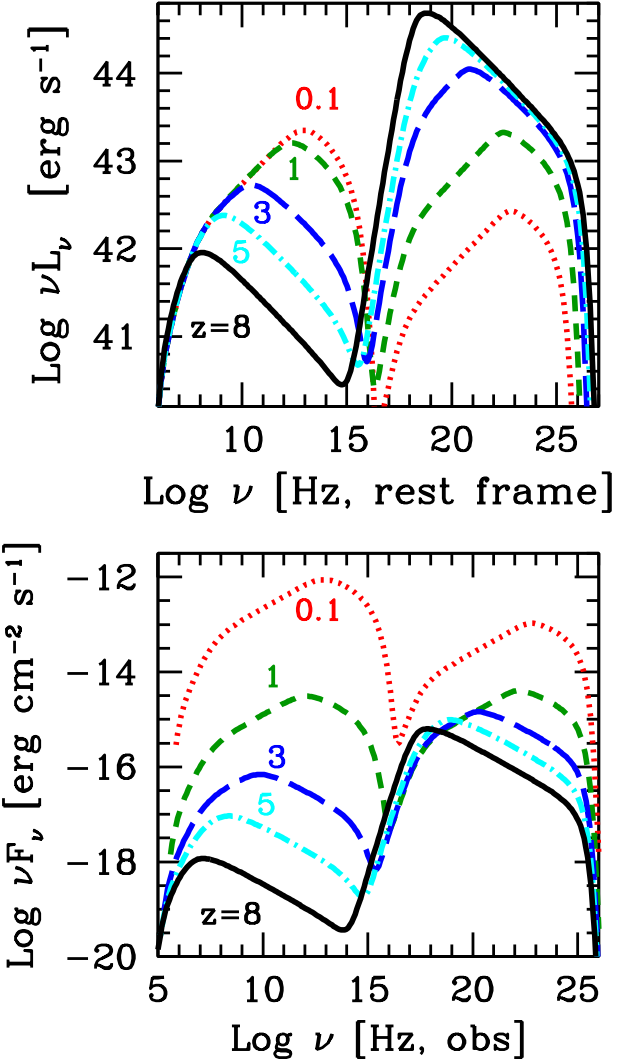,height=16cm,width=16cm} 
\vskip -0.8 cm
\caption{
Top panel: as in Fig. \ref{loboz}, but with a  source size $R\sim 25$ kpc.
The magnetic field is 10 $\mu$G. 
Below the cooling synchrotron frequency, the radio luminosity is
independent of redshift, i.e. it is not affected by the increased  $U_{\rm CMB}$.
At high radio--mm frequencies instead (above the cooling frequencies)
the luminosity decreases as $z$ increases.
Bottom panel: the same  SEDs in the observer frame.
} 
\label{lobor}
\end{figure}

\subsubsection{Changing the source size}

In order to test the role of the source size, i.e. the escape timescale, 
we considered   models where $R=25$ kpc, a quarter of what previously assumed. 
Fig. \ref{lobor} shows the SED for a magnetic field $B=10$ $\mu$G, with all
the other parameters as in the double power--law models shown in
Figs. \ref{loboz}, \ref{lobozf}.
The top panel  illustrates the SEDs in the rest frame of the source, while
the bottom panel reports the same SEDs as observed at $z=0$.
Differently from the $R=100$ kpc case,  we note that in 
a range of (low) radio frequencies the thin synchrotron luminosity is redshift independent 
(see top panel, models from $z=0.1$ to $z\sim 5$),
while the luminosity differs at different redshifts above the peak frequency.
This implies that if two similar sources located at different redshift 
are observed above -- say -- 10 GHz, high $z$ extended source
might be so faint to be undetectable, while observing at -- say --
100 MHz  both would be visible and with similar luminosities. 
This  offers  a clear opportunity to test the CMB effect on
extended radio sources as proposed here, by using, e.g., LOFAR.

\section{Discussion and conclusions}

Beaming makes blazars a unique tool in assessing the number density of radio--loud 
AGN hosting supermassive BH at high redshift.
Indeed, for any confirmed high--redshift blazar there must exist other 
$\sim 2\Gamma^2 =450 (\Gamma/15)^2$ sources sharing the same intrinsic properties, whose jets are 
simply not pointing at us. 
In Ghisellini et al. (2010a) and Volonteri et al. (2011) this idea was exploited to show 
that the comoving number density of supermassive BHs ($M>10^9M_\odot$)
powering {\it Swift}/BAT (Gehrels et al. 2004) blazars peaks at $z\simeq 4$ 
(see also Ajello et al. 2009). 

The obvious question that followed concerned the observable properties of the 
hypothetically large parent population of (misaligned)  radio galaxies at  high redshift.  
Indeed, such large population may be largely missed in existing radio surveys. 
Here we have discussed how the interaction between the electrons and the redshift 
dependent CMB energy density affects the emission of extended structures, 
making the appearance of intrinsically similar radio--loud AGNs different at different cosmic epochs. 
Our findings can be summarized as follows:

\begin{itemize}

\item 
For $z\gsim3-5$ the CMB energy density becomes dominant, and partly suppresses
the synchrotron flux in extended radio sources. 
This  implies that the spectral properties and the statistics of 
radio galaxies at high redshifts could be very different from their low--$z$ counterparts. 

\item 
We do not expect the radio lobes to be homogeneous, in terms of particle density and magnetic field. 
In regions where the magnetic field energy density is larger than the CMB one, $U_{\rm B}>U_{\rm CMB}$,  
the electrons are bound to cool down by synchrotron emission anyway, and hence are still efficient radio--emitters. 
Among such regions  are the so--called hot--spots. 
Therefore the observed contrast in radio emission between compact, magnetized regions 
and more extended lobes should increase for increasing redshifts. 

\item 
At low enough energies, radiative cooling is inefficient even considering
the IC scattering with CMB photons, at all redshifts.
Therefore, the luminosity at the corresponding radio frequencies does 
not depend upon redshifts (all other parameters being equal).
This is potentially testable by observing at high and low radio frequencies.

\item
The decrease of the radio flux due to the  increase of IC off CMB photons could
have important consequences when searching for radio--loud sources 
in large samples, because it introduces an obvious selection effect. 

\item
If the hot spot is visible, but not the extended lobes, 
a classical double radio source can be misclassified as two different
point--like  radio sources at relatively large angular separation. 
As an example, a linear projected distance of 100 kpc is seen
under an angle of 13, 14.3 and 16 arcsec, at $z=$3, 4 and 5, respectively.
Therefore, when cross correlating sources in optical and radio catalogs, 
we could be missing some identifications just because a given optical source
has no radio counterparts within a pre--assigned search radius.

\vskip 0.2 cm

In conclusion, we have shown how enhanced inverse 
Compton cooling of relativistic electron 
off CMB photons at high redshifts can potentially alter the radio and X--ray emissions of 
extended structures associated to radio--loud AGNs. 
This has important consequences for the statistics of distant radio sources. 
Potentially, radio surveys could miss many high--redshift {\it misaligned}
sources because of the reduced radio flux of their extended components.
Instead, in {\it aligned} sources (i.e. in blazars), the observed flux comes predominantly from the 
relatively compact jet, whose radio emission is almost unaffected by the CMB.
Therefore blazars become the most powerful proxies to estimate the distribution and 
number density of the total (i.e. parent) population of radio sources, and to assess
the number density of the associated heavy black holes.

\end{itemize}

\section*{Acknowledgements}
We thank Stefano Borgani for useful discussions. FT thanks financial support
from a PRIN--INAF 2011 grant.

\end{document}